\documentclass[dvips]{acta}
\usepackage{supertabular,lscape,epsfig}
\usepackage{amssymb}
\usepackage{amsmath}
\usepackage{rotating}

\newcommand{\MakeHorTable}[4]{\begin{sidewaystable}[!hp] \begin{center} \TabCap{#2}{#3}
   \TableFont \begin{tabular}{#1} #4
  \end{tabular}\end{center}\end{sidewaystable}}

\SetPages{0}{0}

\SetVol{58}{2008}

\begin{document}

\begin{Titlepage}
\Title{Coronal activity from the $ASAS$ eclipsing binaries}
\Author{Szczygiel$^1$,~D.~M., Socrates$^2$,~A., Paczy{\'n}ski$^2$,~B., Pojma\'nski$^1$,~G., Pilecki$^1$,~B.}
{$^1$Warsaw University Observatory, Al. Ujazdowskie 4, 00-478 Warsaw, Poland\\
e-mail: dszczyg, gp, pilecki@astrouw.edu.pl\\
$^2$Princeton University Observatory, Peyton Hall, Princeton, NJ 08544, USA\\
e-mail: socrates@astro.princeton.edu}

\Received{Month Day, Year}
\end{Titlepage}

\Abstract{
We combine the catalogue of eclipsing binaries from the {\it All Sky
Automated Survey} ({\it ASAS}) with the {\it ROSAT} All Sky Survey
({\it RASS}).  The combination results in 836 eclipsing binaries that
display coronal activity and is the largest sample of active binary
stars assembled to date.
By using the ({\it V-I} ) colors of the {\it ASAS} eclipsing binary
catalogue, we are able to determine the
distances and thus bolometric luminosities for the majority of
eclipsing binaries that display significant stellar activity.  A
typical value for the ratio of soft X-ray to bolometric luminosity is
$L_X/L_{\rm bol}\sim$ a few $\times 10^{-4}$, similar to the ratio of
soft X-ray to bolometric flux $F_X/F_{\rm bol}$ in the most
active regions of the Sun.  Unlike rapidly rotating
isolated late-type dwarfs -- stars with significant outer convection
zones -- a tight correlation between Rossby number and activity of
eclipsing binaries is absent. We find evidence for the saturation effect
and marginal evidence for the so-called ``super-saturation'' phenomena.
Our work shows that wide-field stellar variability searches can
produce a high yield of binary stars with strong coronal activity.

The combined {\it ASAS} and {\it RASS} catalogue, as well as the results
of this work are available for download in a form of a file.}

{stars: eclipsing -- stars: binary -- stars: evolution -- stars: X-rays}

\section{Introduction}

In the last few decades, it has become clear that chromospheric and
coronal activity of main sequence dwarfs is intimately tied to stellar
rotation rate.  Main sequence stars are born rapidly rotating
and quickly lose their angular momentum by some combination of stellar
winds and magnetic torques.  It follows that stars with rapid rotation
-- ideal for studies of stellar activity -- are quite rare.

In comparison to isolated main sequence stars, close eclipsing
binaries are rare as well.  However, they are dramatically variable
in comparison to rapidly rotating single stars.
Close binaries that consist of two late-type dwarfs
are thought to be synchronized as long as their orbital periods are
shorter than $10$ days or so (Duquennoy \& Mayor, 1991).
This is true for the majority of close binaries, which then may be
fruitfully employed in a study of stellar activity.

Contact binary stars were discovered to be strong X-ray emitters almost
30 years ago (Carroll et al. 1980). Shortly afterward Cruddace and
Dupree (1984) identified 14 X-ray active W UMa systems and noticed
a correlation between X-ray luminosity and rotational period,
suggesting that for very fast rotators this relation may be reversed.
Since then, the study of W UMa activity has broadened -- X-ray flux
variations were discovered (McGale et al. 1996, Brickhouse
and Dupree 1998), the color-activity and period-activity dependencies
were studied on a bigger sample of 57 W UMa binaries
(St\c epie\'n et al. 2001). But there is still no final theory
describing the mechanism of X-ray emission.

Recently large databases of contact binaries were searched for coronal
activity, offering still larger binary samples and thus better
statistics. Geske et al. (2006) combined the Northern Sky Variability
Survey (NSVS) catalogue with the ROSAT All-Sky Survey X-Ray catalogue
and found 140 new binaries with active corona.
A part of the All Sky Automated Survey (ASAS) catalogue (southern
hemisphere) was also investigated (Chen et al. 2006), and 34 objects
more were identified. In 2005 ASAS released the complete catalogue of
variable stars south of declination +28deg (Pojma\'nski et al. 2005),
among which there are over 11,000 eclipsing binaries, thus providing
the largest sample of galactic field binary stars.

In this short observational paper we combine the complete ASAS catalogue
of variable stars (ACVS) with the ROSAT All-Sky Survey X-ray catalogue
(RASS). We obtain the largest sample of X-ray active contact binary
stars up to date, which constitutes 379 sources. Furthermore, we widen
the study of stellar activity to include all types of eclipsing binaries
i.e., including semi-detached and detached as well.
The overall number of coincident sources amounts to 836.
The overwhelming majority of the objects in our sample are close binaries
that are expected to be in tidal synchronization.
Therefore, we take the orbital period to be equal to the
rotation period of each individual star.

Sections 2 and 3 describe both the ASAS and ROSAT catalogues, the process
of preparing the combined sample and incidence of X-ray activity.
In Section 4 we discuss the method of determining the bolometric
and X-ray luminosities. The complete catalogue is presented in Section 5.
A short analysis of the evolution of X-ray, or coronal, activity
with rotation and colour is given in Section 6. We summarize the paper
in Section 7.

\section{The ASAS eclipsing binary catalogue }

The All Sky Automated Survey (ASAS) is a blind optical survey that
covers approximately 3/4 of the night sky.  The scientific goal of
ASAS is to study variability at the bright end ($8<V<14$ mag), and at
a moderate cadence (1-3 days).  For the last 7 years, ASAS has been
collecting data in both the $V$- and $I$- band.  At this sampling rate
and limiting magnitude, ASAS is optimized for studies of stellar
variability.  To date, ASAS has generated a rich {\it V}-band
catalogue of over 50,000 classified
variable stars.\footnote{Information on ASAS
(Pojma\'nski 1997, 1998, 2000, 2002, 2003, Pojma\'nski and Maciejewski
2004, 2005, Pojma\'nski, Pilecki and Szczygie\l{} 2005) and its freely
accessible data are located at\newline
http://www.astrouw.edu.pl/asas\newline
The reduced data are available in both download-able ASCII format as
well as an on-line database.}

Roughly 20\% (11,076) of the ASAS variable stars have been identified
as eclipsing binaries (Pojma\'nski 2002; Paczy\'nski et al. 2006).
The ASAS eclipsing
binaries are divided into three classes, which are referred to as
Eclipsing Contact (EC), Eclipsing Semi-Detached (ESD) and Eclipsing
Detached (ED).  Roughly 50\% of the ASAS eclipsing binaries from the
catalogue belong to the EC group while the other half is approximately
evenly split between ESD and ED.

Note that the classification scheme of Pojma\'nski (2002), that we
employ here,
differs from the more traditional convention of dividing eclipsing
binaries into EW, EB and EA types. The (EC, ESD, ED) scheme
discriminates between eclipsing binaries based upon the photometric
smoothness of the eclipse.  In the case of EC binaries, each eclipse
is relatively smooth, while each eclipse is relatively sharp for the
ED binaries.  Therefore, the ratio of primary radius to orbital
separation is relatively large for the EC systems while for ED
binaries, this ratio is relatively small.  A strong motivation for
utilizing the Pojma\'nski (2002) convention is the
abundance of photometric data from ASAS as well as the (initial)
lack of spectroscopic information for the ASAS variable star
catalogue.

In this work, we take advantage of the {\it I}-band filter of ASAS,
whose reduced photometric observations will soon be publicly available.
As a result, we are able to utilize both photometric and color
information of the 11,076 ASAS eclipsing binaries.

\section{Combining the ASAS eclipsing binaries with ROSAT}

\begin{figure}
\begin{center}
\includegraphics[width=0.48\linewidth]{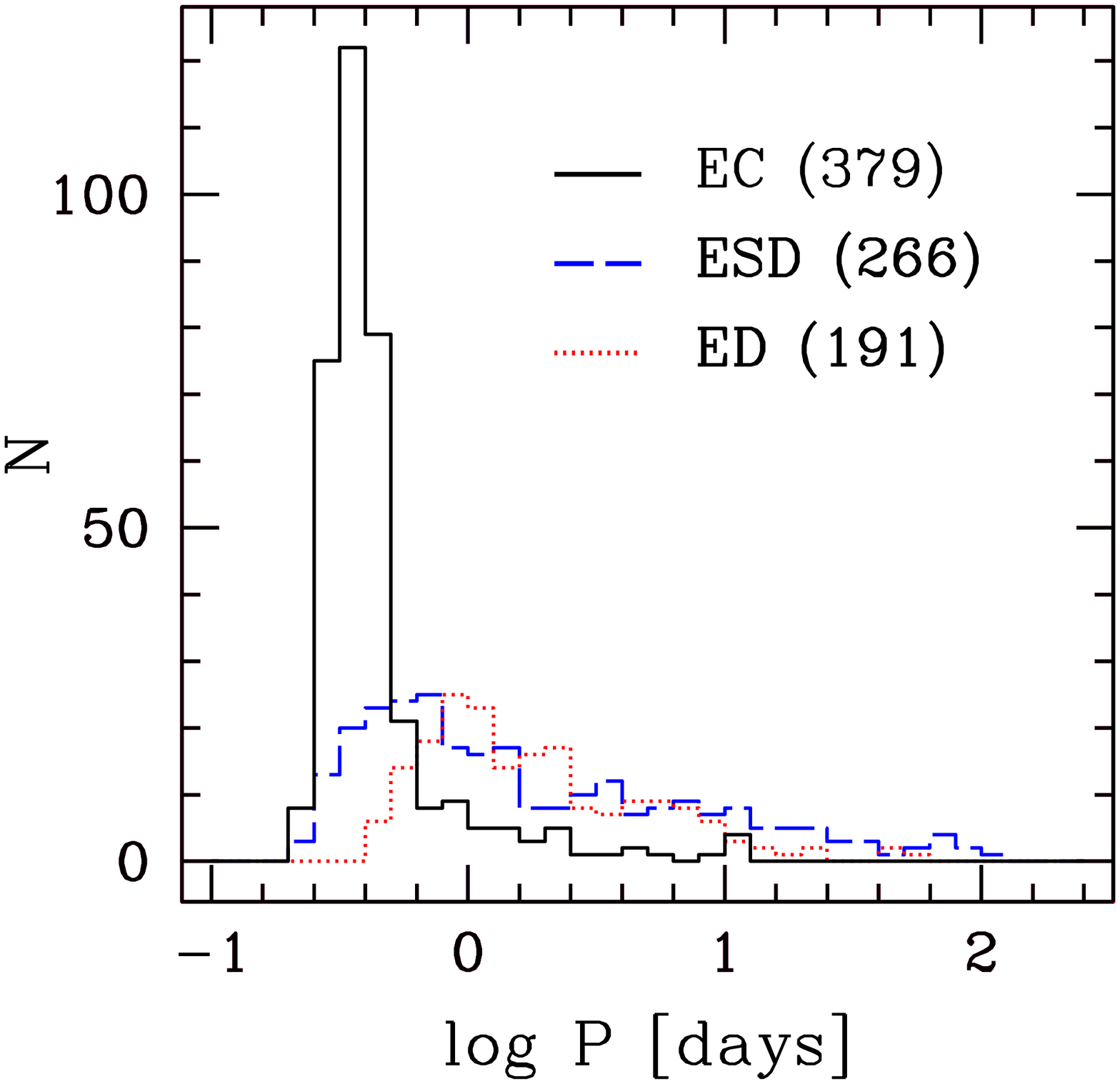}
\end{center}
\FigCap{Orbital period distribution of the ASAS eclipsing binaries
that were matched with X-ray sources in the RASS catalogue.}
\end{figure}

The soft X-ray components of stellar spectra are primarily thought to
originate from coronal activity.  Stars on the main
sequence that possess sizable outer convection zones exhibit
increasing levels of coronal activity with increasing rotation rate.  In
particular, the ratio of X-ray to bolometric luminosity $L_X/L_{\rm bol}$
for G-F stars increases in proportion to the inverse of the Rossby number
${\rm Ro}=P/ \tau_{c}$ squared (Schmitt et al. 1985), where $P$ and
$\tau_{c}$ are the orbital period and convective turnover time of the
primary, respectively. This relation holds as long as ${\rm Ro} \geq 1$.

As is readily shown in Fig.~1, an overwhelming majority
of the ASAS eclipsing binaries possess orbital periods shorter
than ten days (which is purely a result of observational selection) and
hence, are tidally locked.  It follows that most of the individual stars
in Fig.~1 are rapidly rotating.

In order to study the relationship between coronal activity, rotation,
and spectral type in the ASAS eclipsing binaries, a relatively
deep wide-field X-ray survey is required.  For these purposes,
the best data set to date is given by the ROSAT All Sky Survey
(RASS), which we briefly describe below.

\subsection{Combined ROSAT and ASAS EB catalogue}

Data for the RASS were taken primarily between 1990 and 1991 and holes
in the survey were filled with pointed observations in 1997.  The sky
was scanned in 2$^{\circ}$ strips along the ecliptic and therefore
regions of low ecliptic latitude were more thoroughly surveyed.  The
RASS covers between 0.1-2.4 keV in photon energy and entries are
divided into a bright and faint source catalogue, consisting of 18,811
and 105,924 entries, respectively.  Details on the ROSAT satellite
and the RASS can be found in Voges et al. (1999).

We combine the ASAS eclipsing binary catalogue with the RASS Bright
and Faint Source catalogues.  The angular resolution of
ASAS and RASS is $\sim 15''$/pixel and $\sim 50''$/pixel,
respectively.  The ROSAT angular resolution of $=50''$
determines whether or not a ROSAT source
and a given ASAS eclipsing binary coincide with one another.
In other words, as long as the angular separation between an ASAS
eclipsing binary and an entry in RASS is less than $50''$, we
assume that they are the same object.  The distribution
of angular separations for the coincident RASS and ASAS
eclipsing binary sources are given in Fig.~2 (left panel).

The combination of both catalogues produces 836 matches.  The
majority of matches (525) refer to the RASS faint source catalogue,
while the remainder (311) to the bright source catalogue.  The
angular distribution of the ASAS-RASS coincident sources is plotted in
the right panel of Fig.~2.
For angular radius of $50''$, the possibility that our matches
are polluted by background X-ray sources may draw concern.

We construct a false catalogue of eclipsing binaries by randomly
shifting the right ascension and declination of each of the ASAS
eclipsing binaries in between $+5$ and $-5$ degrees, similarly to the
test performed by Geske et al. (2006).  The resulting combination of
the false eclipsing binary catalogue and RASS indicates the level of
random coincidence.  From this exercise, we find that only $1.4\%$
of the matches between the real ASAS eclipsing binary catalogue and RASS
are false at a separation of $50''.$

\begin{figure}
\begin{center} 
\begin{tabular}{c c}
\includegraphics[width=0.48\linewidth]{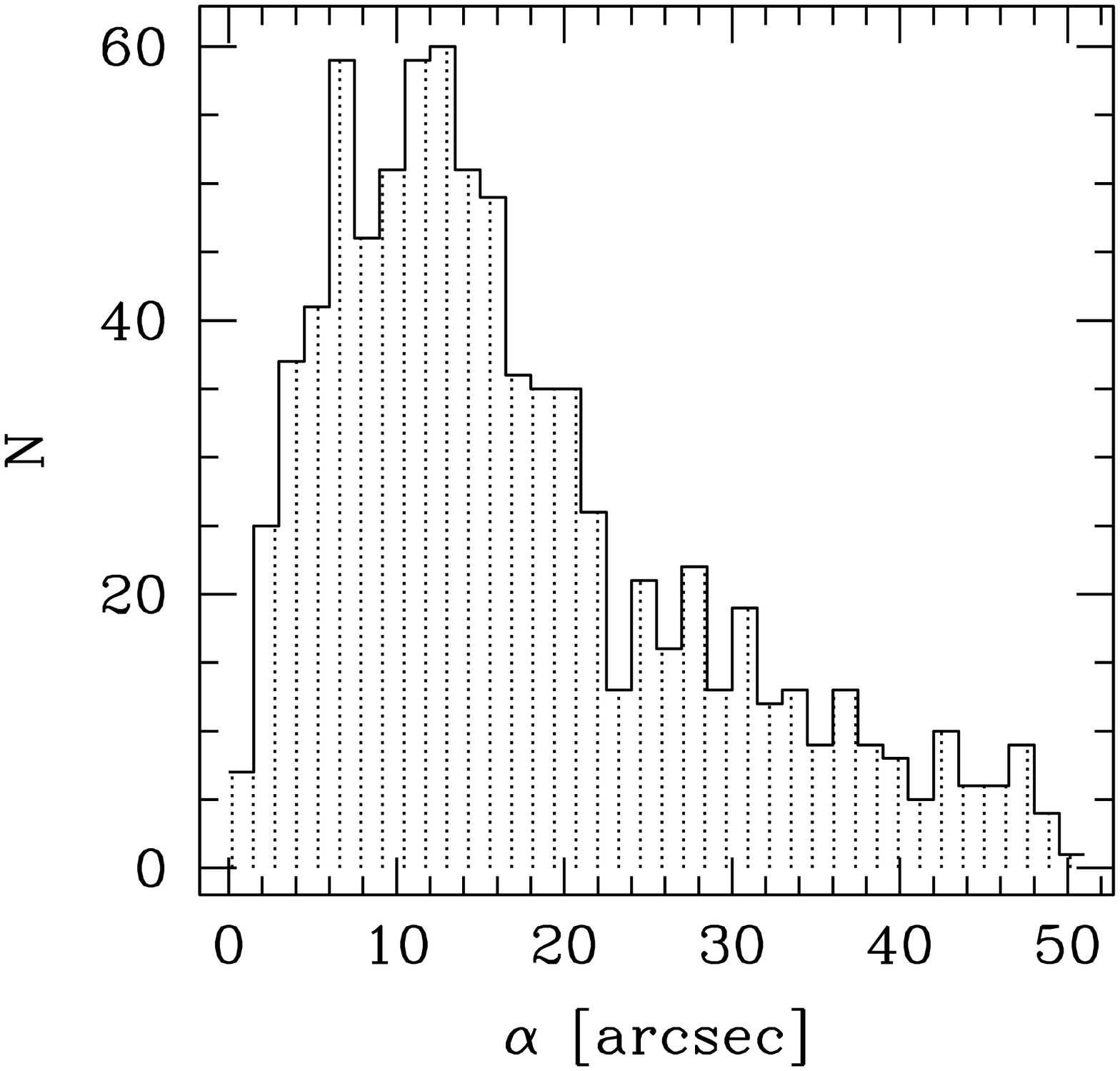} &
\includegraphics[width=0.48\linewidth]{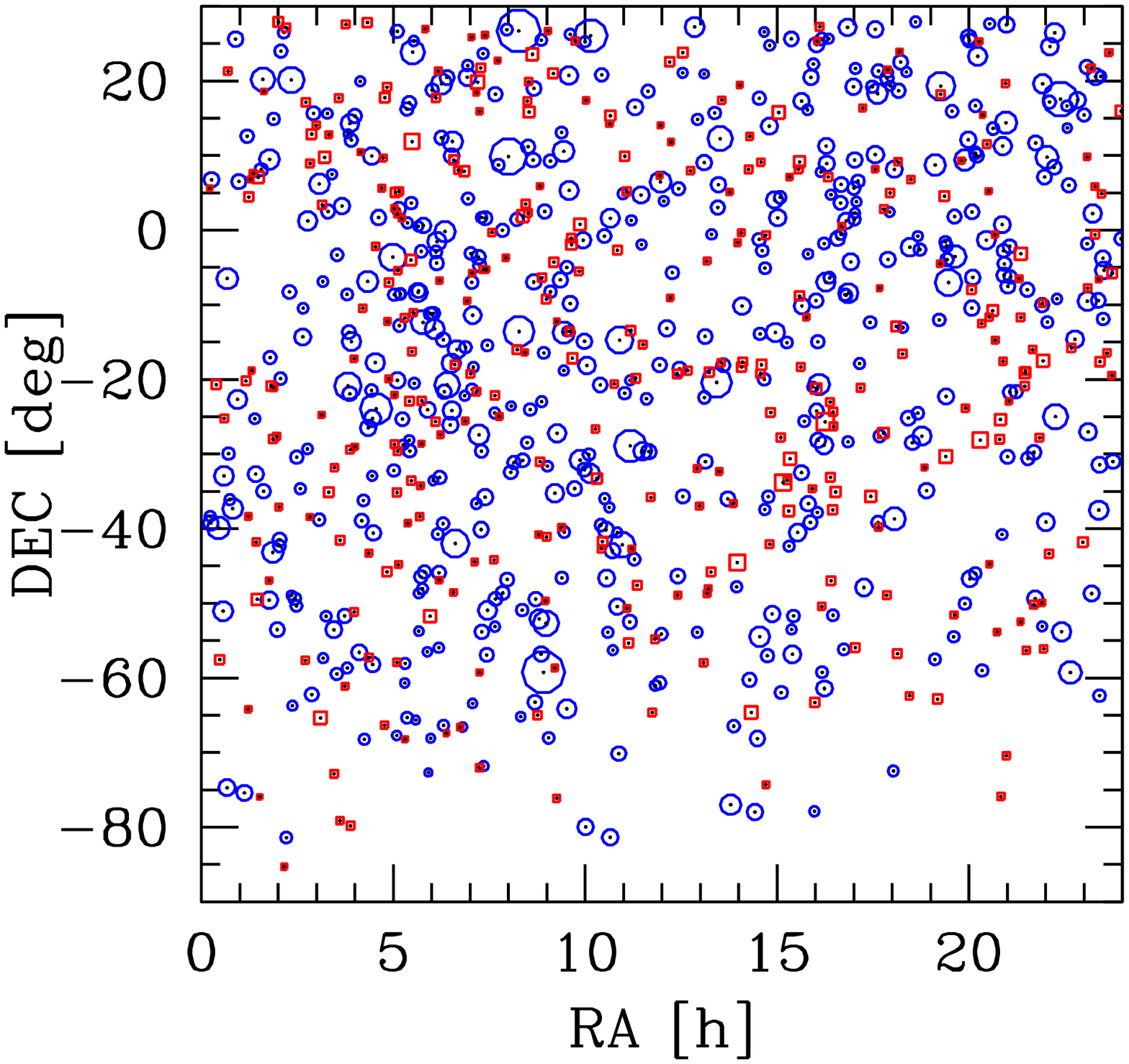} \\
\end{tabular}
\end{center}
\FigCap{Angular separation between matched pairs of the ASAS
eclipsing binaries and the RASS X-ray sources is presented on 
the left panel.
The right panel shows a distribution of objects in combined catalogues
in equatorial coordinates. Blue circles correspond to ROSAT faint
source catalogue, while red squares to ROSAT bright source catalogue.
Symbol size corresponds to positional error in RASS catalogue.}
\end{figure}

As a further check, we loosen the criteria for determining a match
between the catalogues from an angular separation of 50'' up to 90''.
For the latter separation, we find that the number of matches between
both catalogues increases to 942, where 615 and 327 belong 
to the faint and bright source catalogue, respectively.  Thus, 
the increase in number of coincidences is largely due to sources
belonging to the faint source catalogue.  Though we increase  
the number of matches by $\sim 10\%$ by relaxing our criteria for
coincidence, the number of false matches increases to $\sim 3.5\%$.
Nevertheless, from here on, we restrict our study to more
conservative sample, which corresponds to a maximum angular 
separation of $50''$ between RASS and ASAS eclipsing binary catalogues.

\subsection{Incidence of X-ray Activity}

The majority, if not all, of W UMa type variables are thought to be
strong X-ray emitters. 
Cruddace and Dupree (1984) observed 17 W UMa stars, of which 14 showed
X-ray activity. In the sample of St\c epie\'n et al. (2001) 57 out of
100 W UMa were X-ray sources (26 out of 31 closer than 100 pc).
In the work of Geske et al. (2006) 140 X-ray sources were identified
among 1022 galactic contact binaries, which constitutes ~14\%.
However, when the sample is distance limited the percentage is higher
- about 60\% objects within 200 pc, 77\% within 150 pc and 94\% within
125 pc have detectable X-ray emission.

In our distance unlimited sample of 5,376 EC stars 379 exhibit X-ray
activity, which constitutes ~7\%.
We do not have distance estimates for objects which are not coronally
active, so we cannot estimate the change of detection rate with
distance. However, since the ASAS and NSVS projects are similar in the
means of instrumentation and thus sky resolution and magnitude range,
both detection rates should behave similarly with distance.
We assume that roughly 80\% of contact binaries closer that 150 pc
display X-ray activity and in the following studies we will take a look
both at the full sample of active binaries and at this closer subsample.

Since we did not limit our study to contact binaries only, we can
estimate the incidence of X-ray activity among semi-detached and
detached binaries. The former group contains ~9\% active sources
(266 out of 2,957) and the latter ~7\% (191 out of 2,758). These
values are close to the one for EC group, which suggests that there
is a similar number of active binaries of each type.

\section{Fluxes and Luminosities in the Optical and X-ray}

As previously mentioned, there exists a relatively tight
relationship between the stellar activity and the Rossby number
for main sequence stars with convective envelopes.  The resulting
picture is that of magnetic dynamo processes converting
convective and rotational energy  into magnetic energy.
Since magnetic structures are relatively buoyant in comparison to
their surroundings, they have a tendency to rise to the surface, away
from the region of their birth.  The magnetic energy is dissipated
high up in the atmosphere, above the photosphere,  where the effective
resistivity becomes large at relatively low densities.  It is not
clear how precisely this picture applies to close binary systems.

The combination of the ASAS eclipsing binary catalogue with
RASS provides a sizable sample  (836 sources) of binary stars
that display some level of coronal activity.  However, understanding
of the relationship between the spectral type and rotation is less
straightforward for binary systems than for single stars.  Since
a binary consists of two stellar components, it is not clear which
of the individual components is responsible for
the bulk of the stellar activity, though is it clear which component
dominates the total stellar emission.  For example, the
{\it V-}band magnitude of any of the ASAS eclipsing binaries
that we refer to is in reality, the {\it V-}band magnitude at the
maximum of the light curve.  Therefore, the magnitude and color
of the ASAS eclipsing binaries plotted in this paper are
indicative of both components (but mostly of the primary), though
either object may be dominant in X-rays.

For contact binaries the level of uncertainty is even greater.
Neither of the components can truly be thought of as a main sequence
stars from the perspective of a color-magnitude diagram.  That is, a
theoretical understanding of the relationship between surface
temperature, effective gravity, and mass is not well understood for
contact binaries.

\subsection{Bolometric Luminosity of Eclipsing Binaries}

We parametrize the bolometric luminosity $L_{\rm bol}$ by the absolute
magnitude in the {\it V-}band.
\begin{equation}
L_{\rm bol,\star} = L_{\rm bol,\odot} \times 10^{(M_{\rm bol,\odot}-M_{\rm bol,\star})/2.5}
\end{equation}
\begin{equation}
M_{\rm bol,\star} = M_{\rm V,\star} + BC
\end{equation}

Absolute visual magnitudes ($M_{\rm V}$) are calculated from ({\it V-I} )
color; detached and semi detached systems were treated as single stars
and for them we used a main sequence fit by Hawley et al. (1999),
adopting solar metallicity.
For contact binaries we applied a fit of Ruci\'nski and Duerbeck (1997).
With the help of {\it Hipparcos} parallax measurements, they derived
a useful distance calibration for 40 nearby W UMa (contact) binaries.
Since the authors claim, that the coefficients in the formula have large
and non-Gaussian errors, we calculated distances to each star adopting
two formulae - one utilizing {\it V-}band magnitude, and the other using
{\it I-}band.
We rejected 29 contact objects whose distance estimates differed
by more than 20\%.\\

Bolometric corrections ($BC$) were calculated from a polynomial fit to
the data given by Flower (1996), from the $BC - T_{\rm eff}$ relation,
independent of luminosity classes. Effective temperatures were derived
from the ({\it V-I} ) colour, based on calibration of Ram\'irez and
Mel\'endez (2005).

We place the X-ray active ASAS eclipsing binaries on a color-magnitude
diagram in Fig.~3. 
In this and following figures ``EC best'' (filled circles) denotes
a subsample of the EC binaries, which are unambiguously classified
as EC.
Stars, whose type is uncertain are represented with open circles.
Such object are assigned multiple type in the process of the ASAS
automated classification, e.g. EC/ESD or EC/BCEP/DSCT or EC/RRC
(Pojma\'nski, 2002). It happens usually when a light curve is 
of poor quality (eg. for faint stars), or when the amplitude is
small (eg. low amplitude pulsating stars, single spotted rotating
stars, etc.).

Note that all the EC binaries lie above the solid line representing
the main sequence color-magnitude relation of Hawley et al. (1999)
for local stars.
Nevertheless, most of the EC sample lie close to the fit, with the
majority lying about one magnitude above the main sequence relation.
This is not unusual, since some of the contact binaries are the
systems that have already evolved off the main sequence.
Also, some scatter is probably due to mass ratio differences -- 
similar spread is observed on a color-magnitude diagram for
Hipparcos systems (Fig.~1 of Ruci\'nski and Duerbeck 1997).
The objects lying very far above the MS line might simply not be
contact binaries (which is most probable in case of open circles),
thus they do not obey an adequate color-magnitude relation.

\begin{figure}
\begin{center}
\includegraphics[width=0.48\linewidth]{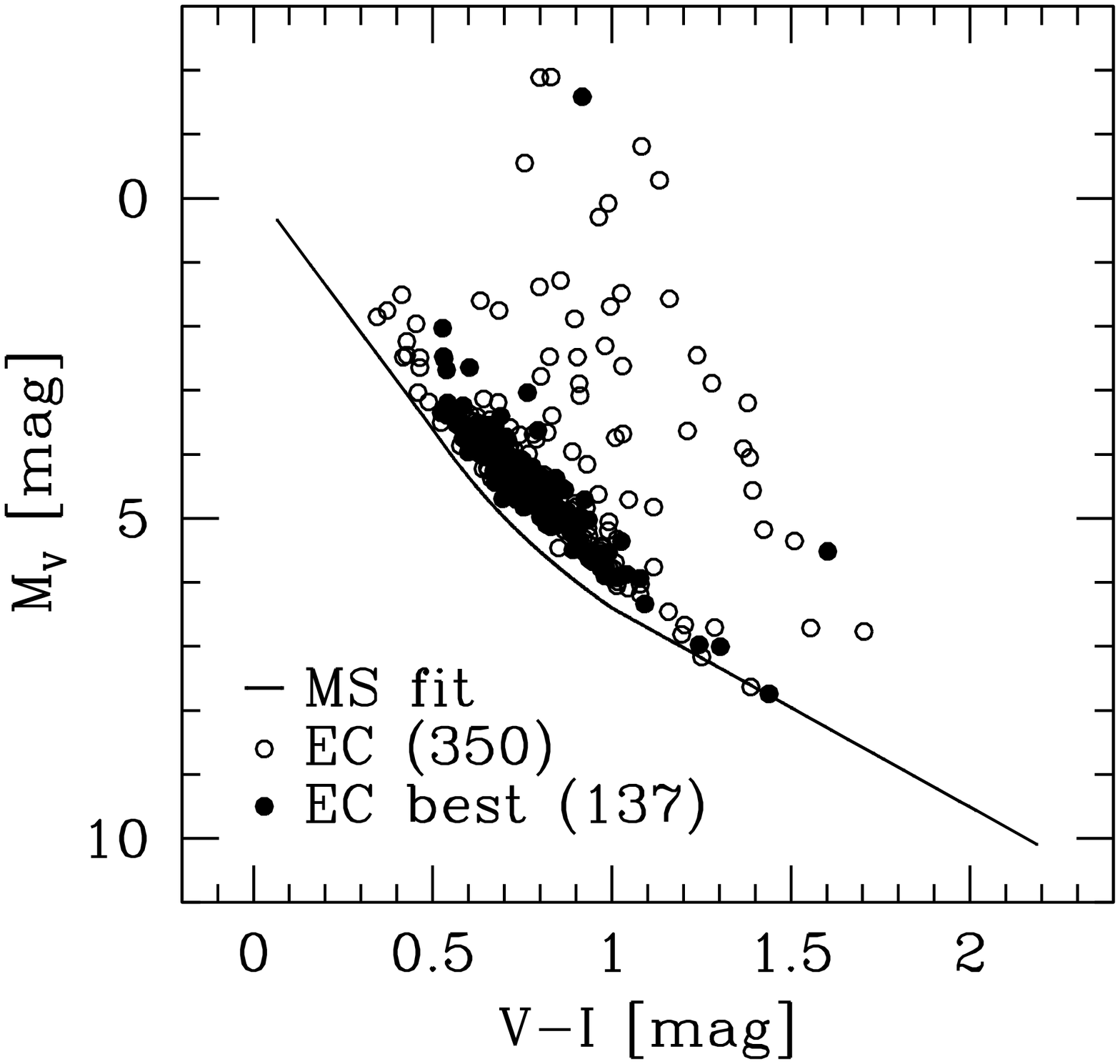}
\end{center}
\FigCap{The colour-magnitude diagram for EC, based on calibration
of Ruci\'nski and Duerbeck (1997). EC best is a subsample of all EC
binaries, in which objects have a unique EC classification (see text).
The line represents a MS relation for single stars derived by Hawley
et al. (1999), which was adopted for ESD and ED variables.}
\end{figure}

For ESD and ED binaries in our sample, we
assume that the luminosity of the system is dominated by the
primary.  Therefore we treat binaries which are either in marginal
contact (ESDs) or detached (EDs) as single stars.  This assumption
may be dangerous when studying a relationship between X-ray emission
and intrinsic physical parameters of the
combined system e.g., the total bolometric luminosity, a size of the
convection zone or orbital period.  First, both stars may not be
in close tidal contact and thus, the rotation period may differ from
the orbital period.  Furthermore, the primary star may not be
a dominant source of the stellar activity.

\begin{figure*}
\begin{center}
\begin{tabular}{c c c}
\includegraphics[scale=.2]{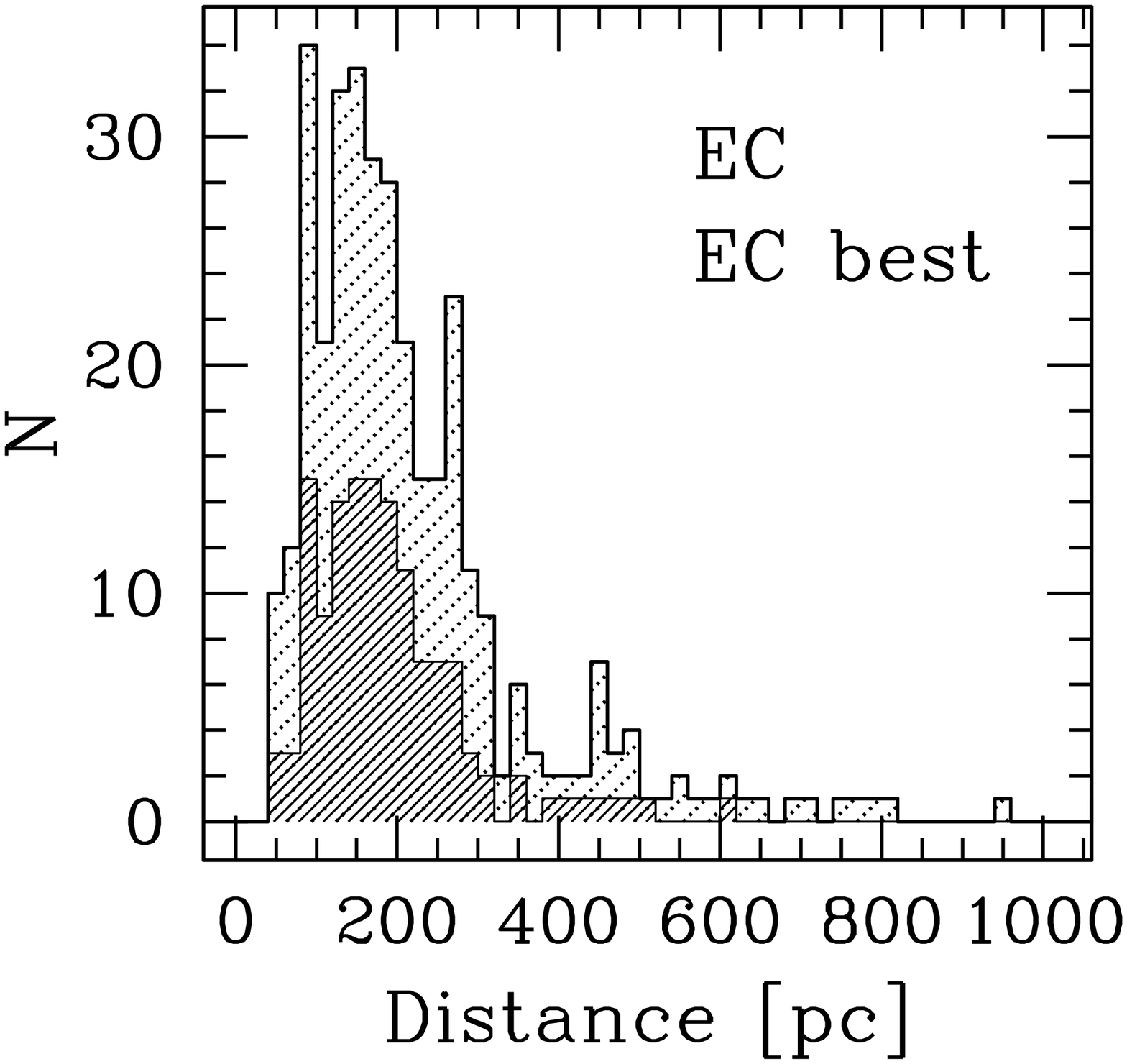}
\includegraphics[scale=.2]{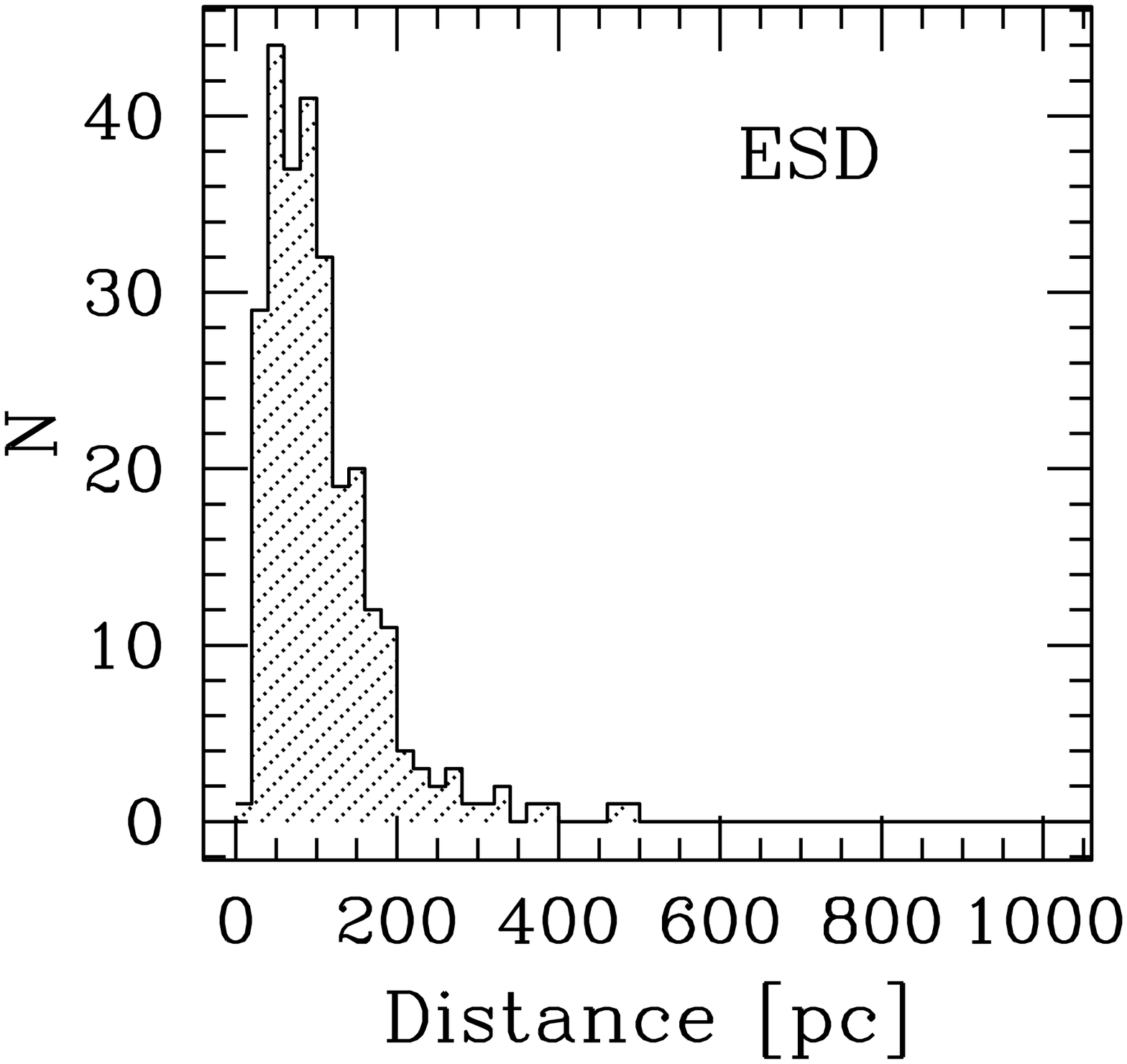}
\includegraphics[scale=.2]{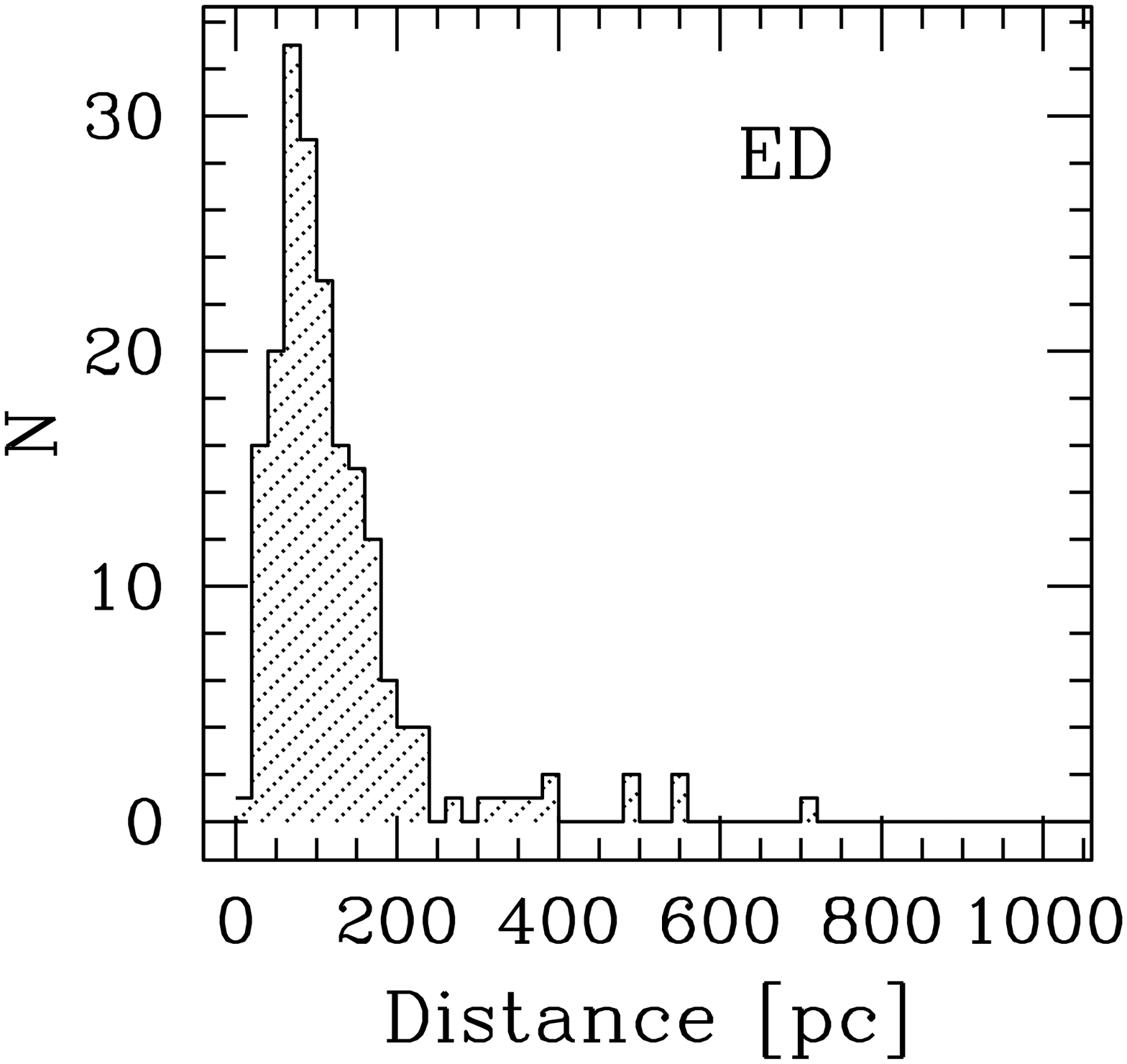}
\end{tabular}
\end{center}
\FigCap{Histograms of estimated distances of contact (EC),
semi-detached (ESD) and detached (ED) eclipsing binaries.
The most left image contains two histograms - the darker one
is based on 'EC best' subsample.}
\end{figure*}

The distance distribution for the three classes of eclipsing
binaries that display coronal activity is displayed in
Fig.~4. 
Systems that are in close contact (EC) are observed to further
distances than semidetached and detached systems (ESD and ED), 
although most of the long tail among ECs is occupied by systems of
multiple type (open circles in Fig.~3).
Since ED are on average more luminous than EC and they are generally
visible to further distances, this may suggest that EC have a higher
level of X-ray activity than ESD and ED.
On the other hand this may simply be a result of the selection effect
- we observe twice as many EC than ED or ESD.

It is important to note that in the whole calculation process described
in this section we did not account for interstellar extinction.
This is justifiable in a close neighbourhood of the Sun, where we
can assume that stars are unreddened, but becomes a significant
factor at larger distances and can introduce a high scatter when not
taken into consideration.
Due to this effect and the fact that the sample completeness
deteriorates quickly with distance (as shown in Section 3.2), we
choose a subsample of active binaries that lie within 150 pc from
the Sun which consists of 127 EC, 214 ESD and 148 ED variables. In
further study we will look both at the full sample and this distance
limited subsample.

\subsection{X-ray Luminosity}

The RASS data provide the number of photons per energy bin per unit
time.  The soft energy bin {\it S} corresponds to photons of energy
0.1-0.4 keV, while the hard bin {\it H} corresponds to photons with
energies of 0.5-2.0 keV.
Huensch et al. (1996) define a hardness ratio {\it HR}
\begin{equation}
HR=\frac{H-S}{H+S},
\end{equation}
which is an important parameter in describing the X-ray spectral
energy distribution.  In terms of the hardness ratio, the X-ray
flux $f_X$ reads
\begin{equation}
f_X=CNTR\times \left(5.30\, HR+8.7 \right)\times 10^{-12} \,{\rm erg\,cm^{-2}\,s^{-1}}.
\end{equation}
where {\it CNTR} is the source count rate.
The spectral energy distribution from which the above relation
originates is based upon {\it ROSAT} observations of a complete,
volume-limited sample of late-type giants (Huensch et al. 1996).
This relation had been previously applied for W UMa type systems,
whose X-ray emitting regions are believed to have similar properties
as those of late-type giants (Chen et al. 2006, St\c epie\'n et al.
2001).\\
Finally, we calculate the X-ray luminosity
\begin{equation}
L_{X} = 4 \pi d^{2} f_{X}
\label{eq_lx}
\end{equation}
of the eclipsing binary in question. The distance $d$ was obtained
from a distance modulus in the {\it V-} band.

\begin{figure}
\begin{center}
\begin{tabular}{c c}
\includegraphics[width=0.48\linewidth]{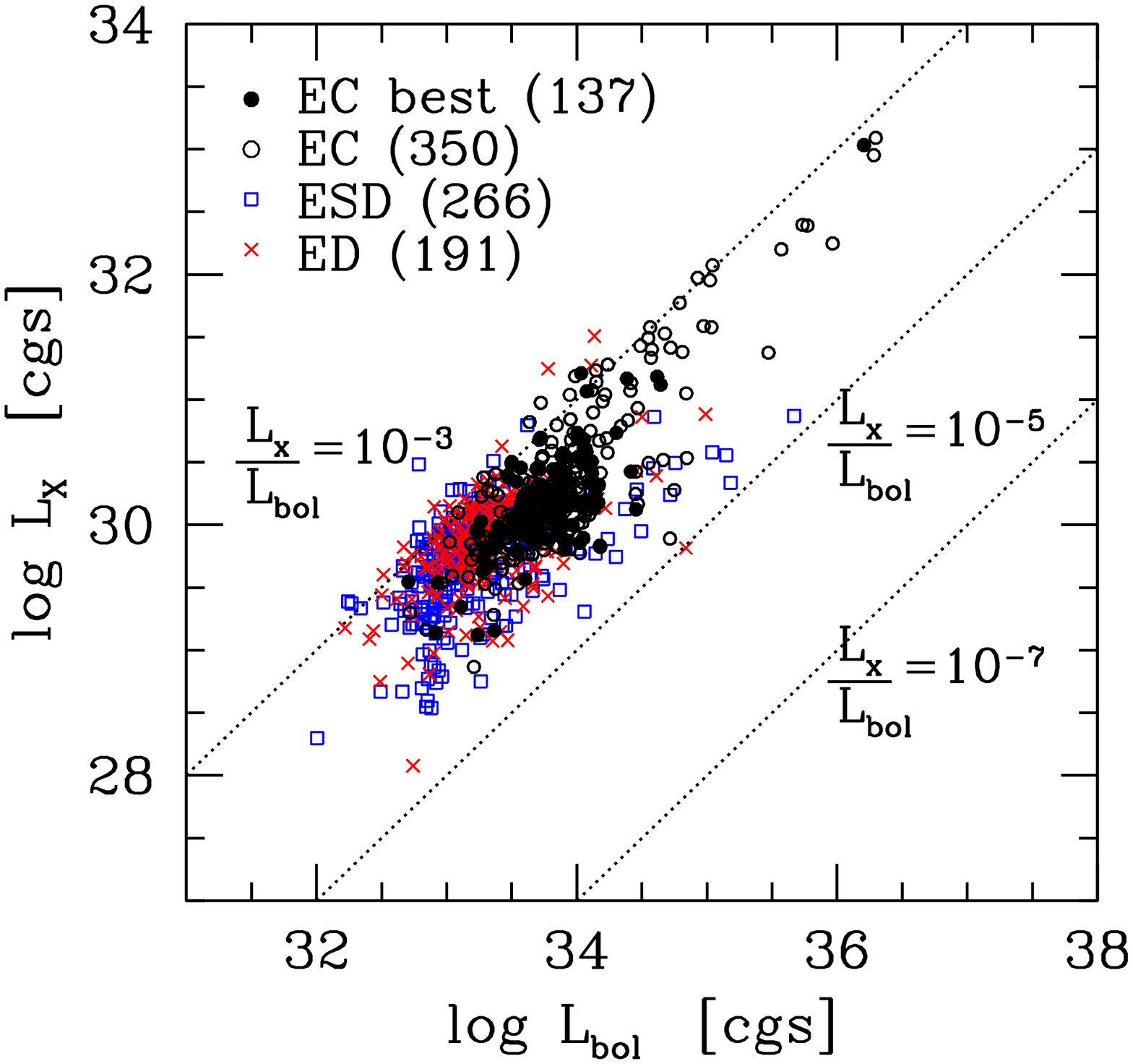} &
\includegraphics[width=0.48\linewidth]{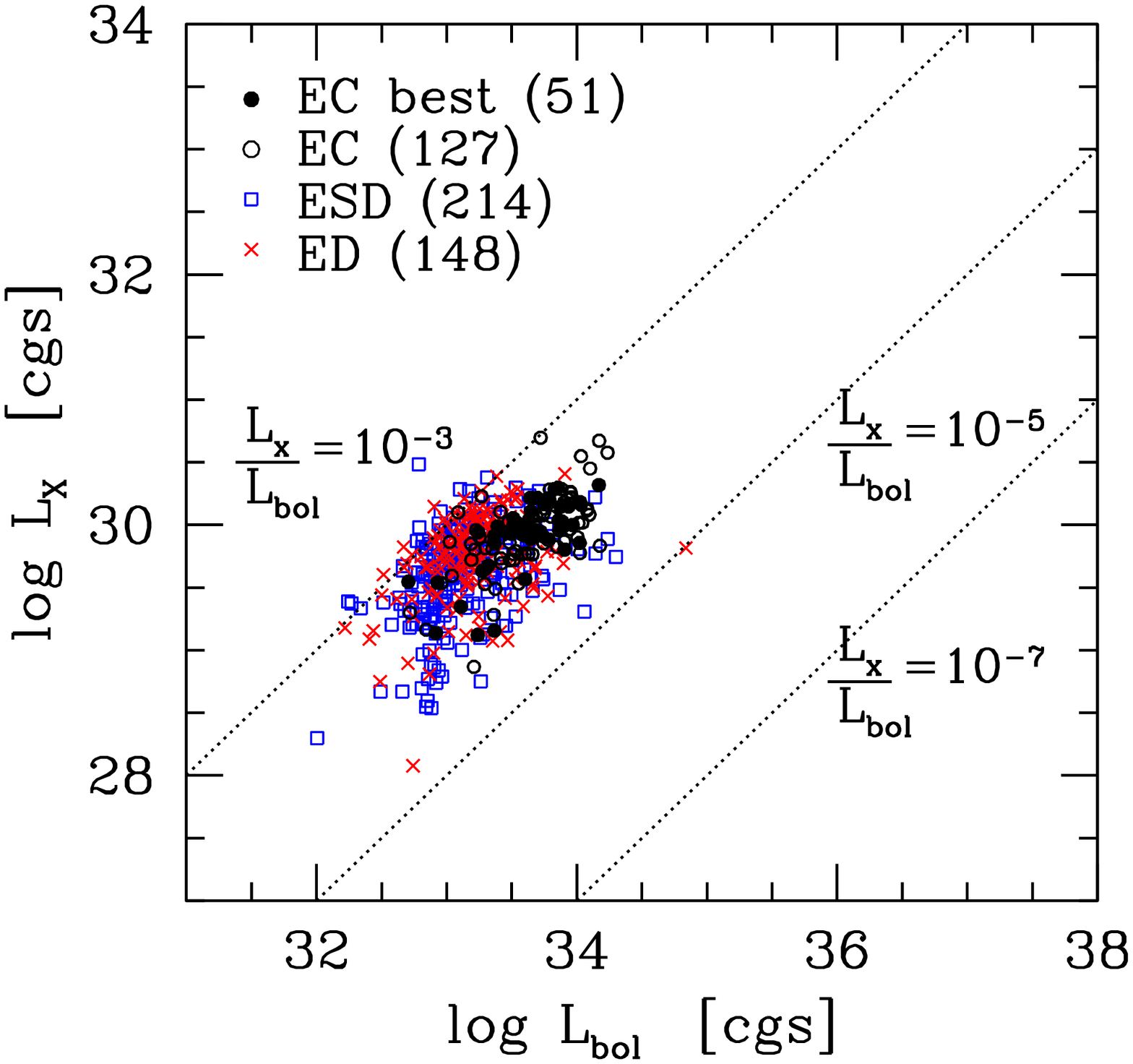} \\
\end{tabular}
\end{center}
\FigCap{Bolometric and X-ray luminosity function of 807 of the
ASAS eclipsing binaries (left) and a distance limited
subsample - $d < 150 pc$ (right).  The overwhelming majority of these
systems lie below $L_X/L_{\rm bol}=10^{-3}$ (dotted line).
EC binaries have higher $L_{\rm bol}$ than ESD and ED groups 
(this is better visible on a colour figure).}
\end{figure}

The X-ray vs. bolometric luminosity for the combined sample is given in
Fig.~5. 
Almost all of the X-ray bright ASAS
eclipsing binaries are active at a level between $L_X/L_{\rm bol}=10^{-3}$
and $L_X/L_{\rm bol}=10^{-4}$
In this sample, and for our choice of bolometric calibration,
EC binaries have higher $L_{\rm bol}$ than ESD and ED binaries,
as was expected following Fig.~3. 

\section{The final catalogue}

\MakeHorTable{ccccccccc}{12.5cm}{ASAS X-ray binaries catalogue.}
{\hline
ASAS ID       &   RASS ID     & $\alpha$ & $P_{orb}$ & $MJD_{0}$ &   V   &   I   & d        & $L_{bol}$ \\
              &               & [arcsec] & [days]    & [days]    & [mag] & [mag] & [pc]     & [erg/s]   \\
\hline
030953-0653.6 & J030952.6-065327 & 10.371 & 0.445286 & 1869.080 & 10.44 &  9.76 & 221.3182 & 34.006 \\
203718-1047.5 & J203718.3-104734 & 5.979  & 0.423600 & 1996.262 & 11.32 & 10.31 &  94.9199 & 33.037 \\
233609-1628.2 & J233608.7-162806 & 7.400  & 6.621000 & 1875.500 & 10.65 &  9.67 &  73.9801 & 33.076 \\
 & & & & & & & & \\
 & & & &.............& & & \\
 & & & & & & & & \\
\hline
.......     & HR  & HR err &  CNTR   & CNTR err & $L_{X}$ & $log(P/\tau_{c})$ & Other ID & Type\\
            &     &        & [cts/s] & [cts/s]  & [erg/s] &                   &          &   \\

\hline
$[table$     & 1.00 & 0.78 & 0.059700 & 0.027530 & 30.690 & -1.422 & UX~Eri & EC\\
$continued]$ & 0.16 & 0.17 & 0.086610 & 0.016190 & 29.950 & -      &   -    & ESD \\
             & 1.00 & 0.29 & 0.064570 & 0.017100 & 29.772 & -0.527 & BQ~Aqr & ED \\

 & & & & & & & & \\
\multicolumn{9}{p{15cm}}{ A few exemplary lines of the ASAS catalogue of X-ray active
binaries ASAS.ROSAT.cat. Each of the 807 lines contains star's ASAS ID
and RASS ID, angular separation of both sources on the sky, orbital
period of the binary, $MJD_{0}$ ($HJD_{0}$ - 2450000), {\it V} and
{\it I} magnitudes at maximum light, distance $d$, bolometric luminosity
$L_{bol}$, hardness ratio and its error, count rate and its error,
X-ray luminosity $L_{X}$ and a logarithm of Rossby number
$log(P/\tau_{c})$. Last two columns contain other ID from the literature
(mostly GCVS) and ACVS variability type (eg. EC, ESD, ED).}
}

We produce a catalog of all ASAS eclipsing binaries for which we
detected \mbox{X-ray} emission.
All data are put in a single file "ASAS.ROSAT.cat" which contains
807 entries. The file is available for download from this website:

\vspace{0.2cm}
http://www.astrouw.edu.pl/asas/?page=rosat
\vspace{0.2cm}
\\Each line in the catalogue contains all basic information about a star,
such as the ASAS and RASS IDs (ie coordinates), magnitudes, orbital
period and all the information from the ACVS that might be useful
for the catalogue users.
Since the RASS catalogues are very large and are easily accessible,
we only include a few quantities from RASS that were used in the course
of this study, that is the hardness ratio and count rate and their
errors. In addition, we supply the catalogue with the values that were
calculated in the course of this study, such as distance, $L_{bol}$,
$L_{X}$ and $log({\rm Ro})$.
The detailed column descriptions as well as a few exemplary lines are
presented in Table 1.

\section{Relationship Between Coronal Activity, Rotation and Color}

Binaries are more complicated than single stars, both on physical
and observational grounds, particularly within the context of stellar
activity.  If indeed there is a physical connection between stellar
activity, convection and rotation as is strongly suggested in the
case of single dwarfs, then a similar trend may exist for binaries
as well.
Nevertheless, it seems difficult to easily disentangle -- from both
an observational and theoretical perspective -- the salient physical
parameters of a given binary for the purpose of quantifying its
activity. Reasons for this difficulty have been previously mentioned.

\begin{figure}
\begin{center}
\begin{tabular}{c c}
\includegraphics[width=0.48\linewidth]{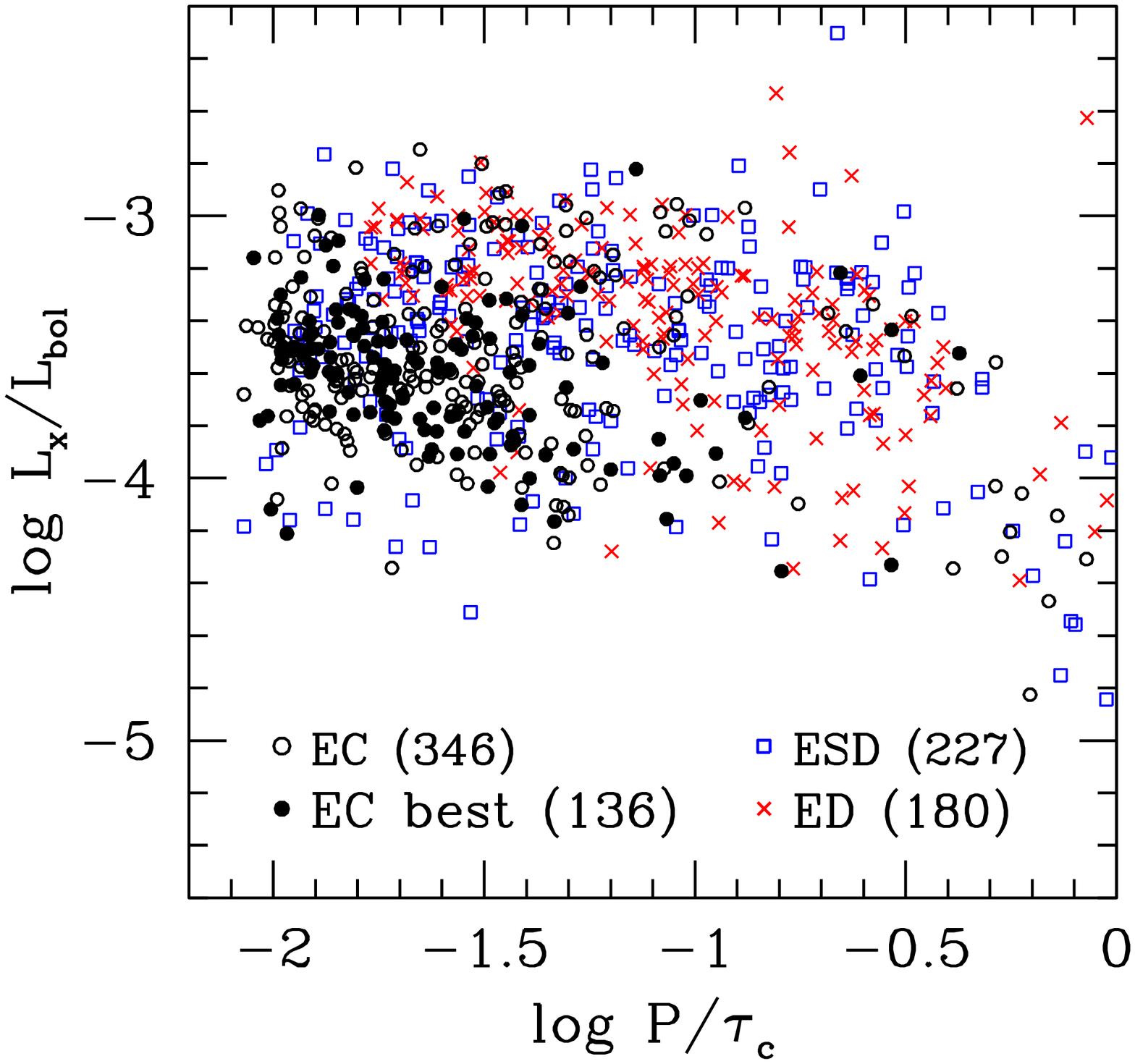}
\includegraphics[width=0.48\linewidth]{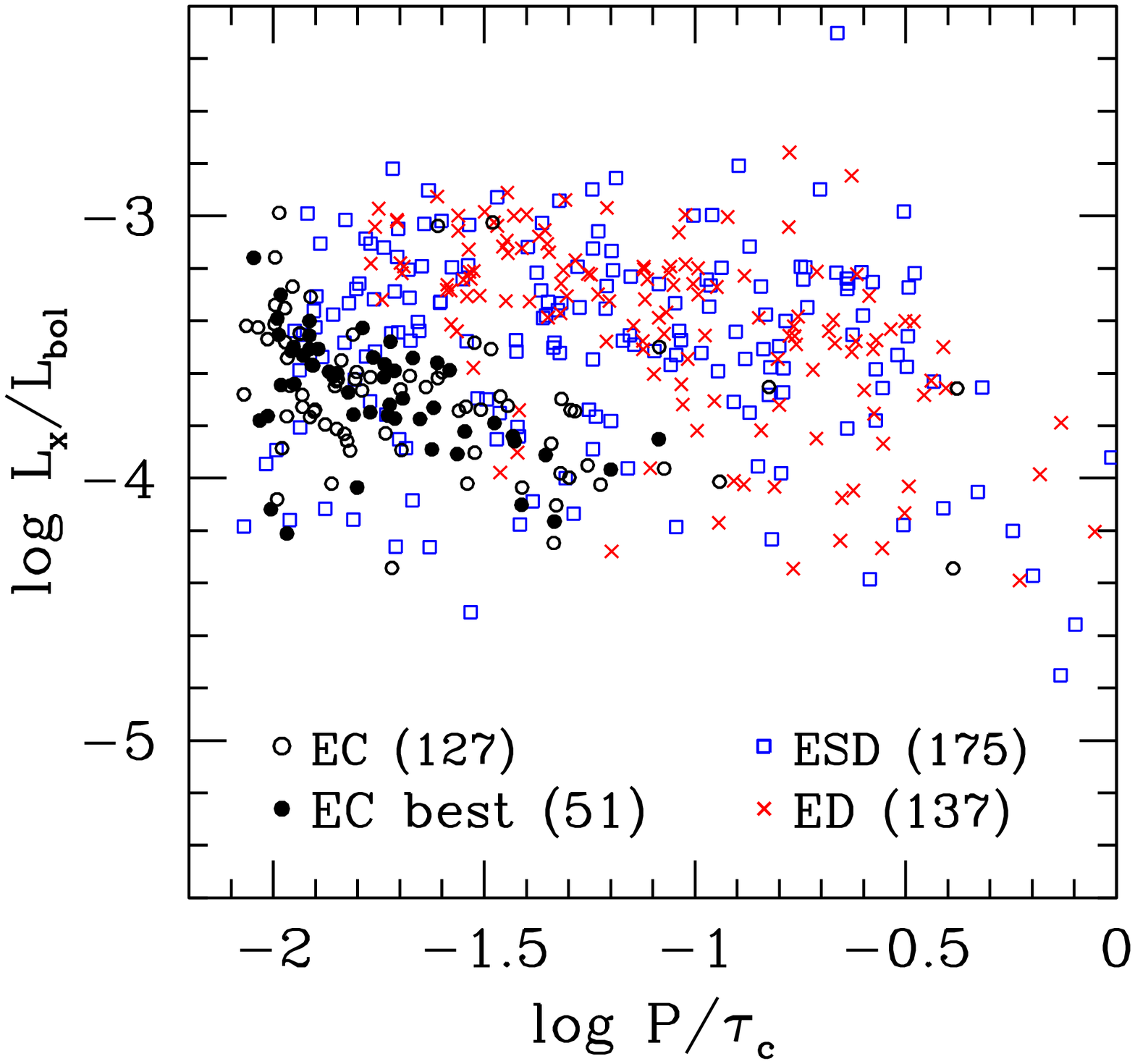}
\end{tabular}
\end{center}
\FigCap{The dependence of activity of the ASAS eclipsing
binaries on the Rossby number, for objects with orbital periods
less than 10 days (753). Left panel contains data points for all
binaries and the right one for objects closer than 150 pc (439).}
\end{figure}

In Fig.~6 
the relationship between activity
and the primary's Rossby number, Ro, is given for the coincident
sources whose orbital periods are less than 10 days, which constitutes
753 objects.  In order
to calculate the Rossby number, the turnover time $\tau_{c}$ for
each star was extracted by the empirical formula provided by
St\c epie\'n (2003).
The turnover time is a function of ({\it B-V}) color, so we transform
our ({\it V-I} ) colors into ({\it B-V})
using the calibration of Caldwell et al. (1993).
The fit of St\c epie\'n is valid for stars with $({\rm \it B-V}) > 0.4$,
so for 34 objects
which had this colour value lower, we adopt $log(\tau_{c})=0$.

Note that relative to the turnover time, all of the binaries in
Fig.~6 
are rapid rotators.  It follows that
the binaries in our sample should be in the "saturated"
state, in analogy with the observed phenomena in rapidly rotating
single late-type dwarfs.  That is, the coronal activity, quantified
by $L_X/L_{\rm bol}$ should be constant with decreasing Rossby number Ro.
The scatter of EC and ESD variables on the activity - Rossby number
diagram is even larger than ECs, and we do not observe a clear increase
of X-ray activity with decreasing rotational period for either group.
We note however, that ESDs and EDs are on average more active than ECs.
To some degree, Fig.~6 
displays the saturation effect in that the level of activity is
roughly constant with decreasing Rossby number.  \\
The same diagram for the limited sample ($d < 150 pc$) does not reduce
the scatter in ESD and ED groups, but the trend of increasing
activity with decreasing Ro is now clearly visible, especially
among contact binaries.

There are claims (St\c epie\'n et al. 2001; Chen et al. 2006)
that the so-called ``super-saturation''
phenomena -- a reduction in coronal and chromospheric activity with
{\it decreasing} Ro -- is observed in extremely rapidly rotating
isolated late type dwarfs {\it and} W UMa binaries.  This subtle
effect is not observed in Fig.~6, 
which is not surprising with the high scatter of points.

\begin{figure}
\begin{center}
\begin{tabular}{c c}
\includegraphics[width=0.48\linewidth]{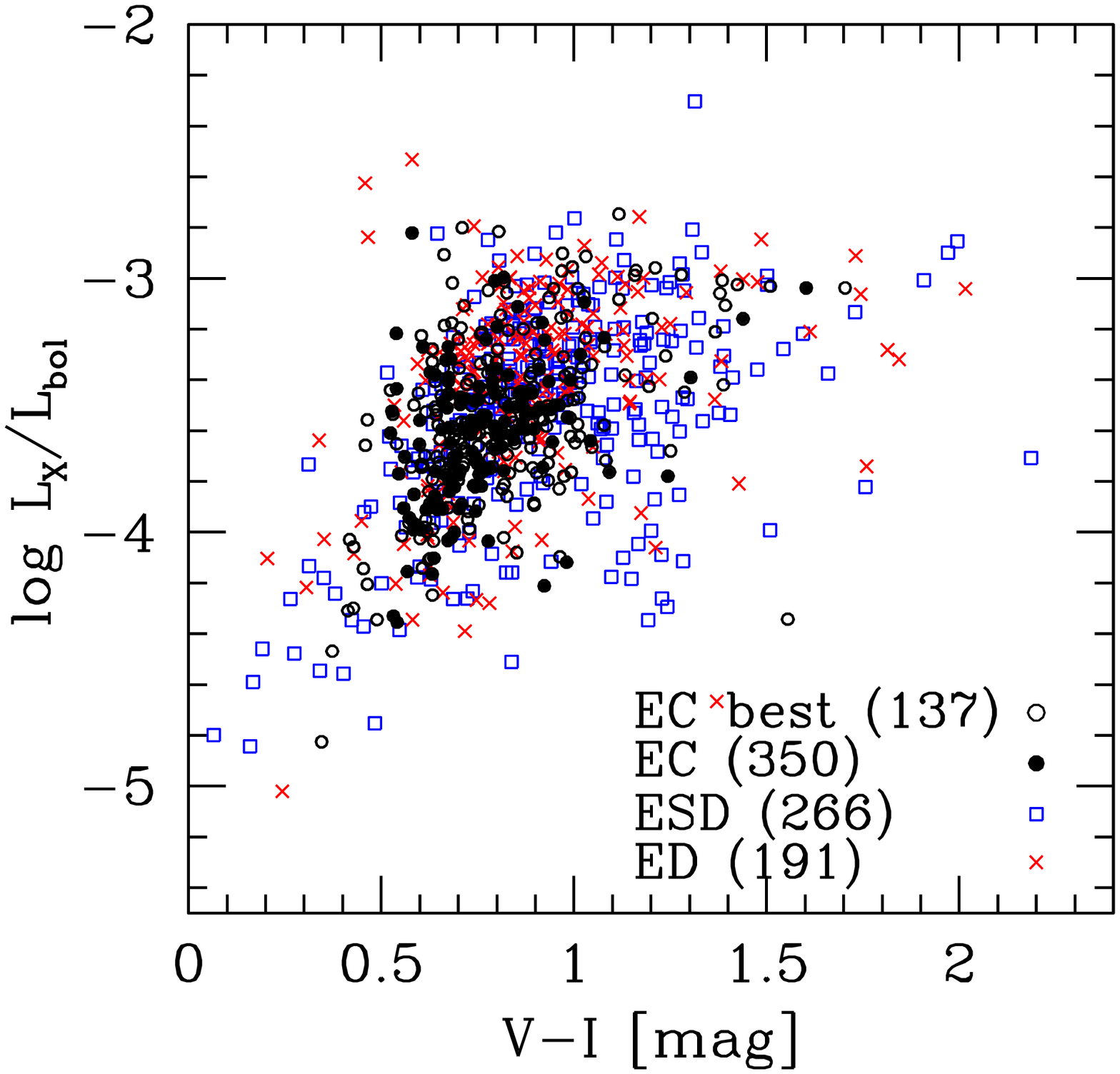}
\includegraphics[width=0.48\linewidth]{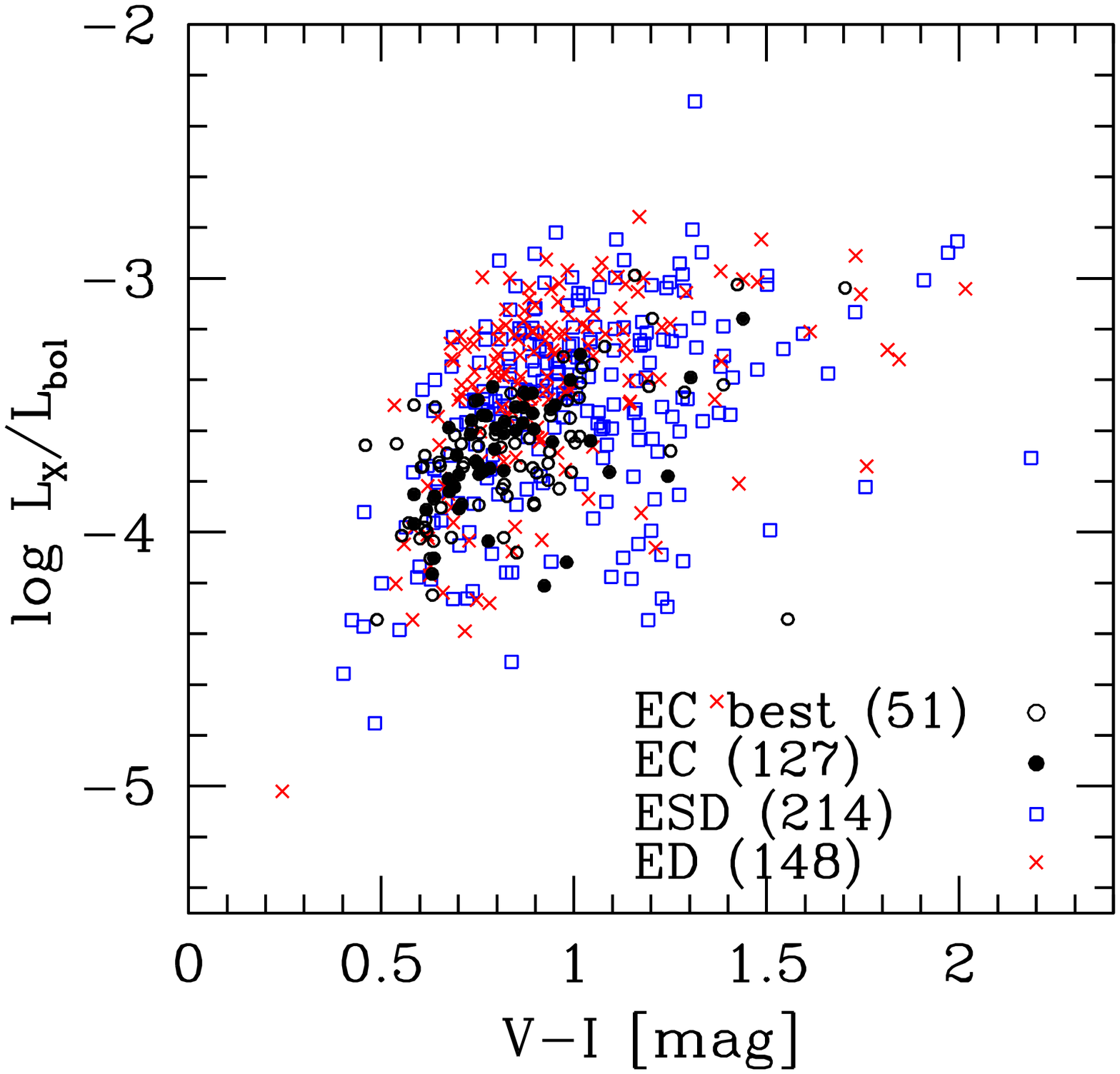}
\end{tabular}
\end{center}
\FigCap{Color-Activity diagram for the ASAS eclipsing binaries
coincident with the RASS. Left panel contains data points for
all 807 binaries and the right one for 489 objects closer than 150 pc.}
\end{figure}

In Fig.~7, 
we plot an activity-color diagram for
all the X-ray bright ASAS eclipsing binaries and for the
subsample. Here, it is clear
that the level of coronal activity indeed changes not only with the
rotation, but surface temperature as well.
With decreasing color index ({\it V-I} ), the surface temperature of
the primary increases, while the depth of the its convection zone
decreases.  So, binaries that are earlier type are less active in
comparison to binaries that are later type.  Of course, in this
context ``early'' and ``late'' refer to surface temperature
alone, rather than surface temperature and rotation as
in the case of isolated stars.

Both Figs.~6 and 7 show that at a fixed overturn time and ({\it V-I} )
colour, the EC binaries are on average less active in comparison
to ESD and ED systems.
This is better visible on diagrams for the subsample of objects closer
than 150 pc.

\section{Summary}

We had produced a catalogue of the combined ASAS eclipsing binaries
and RASS X-ray sources.
All data had been put in a single file "ASAS.ROSAT.cat" which is
available for download from the ASAS website
(http://www.astrouw.edu.pl/asas/?page=rosat) in the form
presented in Table 1 (see Section 5 for details).

The combination of the eclipsing binary catalogue with the RASS
produces a total of 836 coincident sources ($\sim$ a 7\% yield).
Among these are 379 contact binary stars (EC), which is the largest
sample of X-ray active contact binary stars assembled up to date.
Semi-detached (ESD) and detached (ED) binaries were also taken into
account, resulting in 266 and 191 active variables, respectively).
We observe similar incidence of X-ray activity in all 3 groups of
variables, which is around ~7\% in the distance unlimited sample.

An overwhelming majority of the coincident sources possesses orbital
periods shorter than 10 days, so we expect tidal locking between the
primary and the companion.  Therefore, the orbital period of the binary
may be also thought of as the rotation period of the stellar components.

We analyzed the dependence of X-ray activity on Rossby number and
colour. In comparison to typical isolated stars of similar primary mass,
the ASAS eclipsing binaries display a higher level of coronal activity
(see Fig.~5) 
and  the maximum value
$L_X/L_{\rm bol}\sim 3\times 10^{-3}$, for the coincident sources is
similar to that found in rapidly rotating isolated late-type dwarfs.

Both the activity - Rossby number (Fig.~6) 
and the activity - colour (Fig.~7) 
diagrams display a large
scatter and they do not tighten any of the already known relations.
The coronally-active binaries in our sample display the saturation
effect, while there is no clear evidence of the so-called
``super-saturation'' effect.

However, for a given ({\it V-I} ) colour ECs are rotating
more rapidly than the other two classes. And at the same time, for
a given colour the level of activity of ECs is lower in comparison
to ESDs and EDs.
This slight downturn in activity among the three classes may in fact be
an indirect manifestation of the saturation effect in Fig.~7.


\Acknow{We would like to thank K. St\c epie\'n for useful comments
on the paper. This work was supported by the Polish MNiSW grants
N203 007 31/1328 and N N203 304235. AS acknowledges support of
a Hubble Fellowship administered by the Space Telescope Science
Institute and of a Lyman Spitzer Jr. Fellowship at Astrophysical
Sciences at Princeton University.}


\end{document}